\documentclass[a4paper,conference,10pt]{IEEEtran}
\IEEEoverridecommandlockouts
\usepackage{cite}
\usepackage{amsmath,amssymb,amsfonts}
\usepackage{algorithmic}
\usepackage{graphicx}
\usepackage{textcomp}
\usepackage{xcolor}
\usepackage{bbold}
\def\BibTeX{{\rm B\kern-.05em{\sc i\kern-.025em b}\kern-.08em
    T\kern-.1667em\lower.7ex\hbox{E}\kern-.125emX}}

\usepackage{caption}
\usepackage{subcaption}

\usepackage{upgreek}

\usepackage[nowarn,acronyms,nonumberlist,nopostdot,nomain,nogroupskip]{glossaries}
\newacronym{uav}{UAV}{Unmanned Aerial Vehicle}
\newacronym{los}{LoS}{Line of Sight}
\newacronym{csi}{CSI}{Channel State Information}
\newacronym{fanet}{FANET}{Flying Ad-Hoc Network}
\newacronym{manet}{MANET}{Mobile Ad-Hoc Network}
\newacronym{aodv}{AODV}{Ad-Hoc On Demand Distance Vector Routing}
\newacronym{olsr}{OLSR}{Optimized Link State Routing}
\newacronym{smurf}{SMURF}{Stochastic Multipath Routing for FANETs}
\newacronym{basmurf}{BA-SMURF}{Beam Aware Stochastic Multihop Routing for FANETs}
\newacronym{sdn}{SDN}{Software-Defined Networking}
\newacronym{upa}{UPA}{Uniform Planar antenna Array}
\newacronym{dbr}{DBR}{Distance-Based Routing}
\newacronym{sinr}{SINR}{Signal to Interference plus Noise Ratio}
\newacronym{tdma}{TDMA}{Time Division Multiple Access}

\usepackage{pgfplots}
\usepackage{svg}  
\usetikzlibrary{plotmarks}
\usetikzlibrary{arrows.meta}
\usepgfplotslibrary{patchplots}

\pgfplotsset{compat=1.17}

\begin{document}

\title{Beam Aware Stochastic Multihop Routing for Flying Ad-hoc Networks
\thanks{This project has received funding from the European Union’s Horizon 2020 research and innovation program under the Marie Skłodowska-Curie Grant agreement No. 813999. The work of A. Pastore and X. Mestre was supported by Grant RTI2018-099722-B-I00 funded by MCIN/AEI/10.13039/501100011033 and by ``ERDF A way of making Europe''. The author affiliations and emails are as follows:\\ $^1$Department of Information Engineering, University of Padova, Italy. Emails: \{deshpande,zanella\}@dei.unipd.it\\
$^2$Research Unit for Information and Signal Processing for Intelligent Communications (ISPIC),
%Department of Advanced Signal and Information Processing,
Centre Tecnològic Telecomunicacions de Catalunya, Barcelona, Spain. Emails: \{rpereira,apastore,xmestre\}@cttc.es\\
$^3$Department of Electronic Systems, Aalborg University, Denmark. Email: fchi@es.aau.dk\\}}

\author{\IEEEauthorblockN{Anay Ajit Deshpande$^1$, Roberto Pereira$^2$, Federico Chiariotti$^3$,\\ Adriano Pastore$^2$, Xavier Mestre$^2$, and Andrea Zanella$^1$}
}

\maketitle

\begin{abstract}
Routing is a crucial component in the design of \glspl{fanet}. State of the art routing solutions exploit the position of \glspl{uav} and their mobility information to determine the existence of links between them, but this information is often unreliable, as the topology of \glspl{fanet} can change quickly and unpredictably. In order to improve the tracking performance, the uncertainty introduced by imperfect measurements and tracking algorithms needs to be accounted for in the routing. Another important element to consider is beamforming, which can reduce interference, but requires accurate channel and position information to work.
In this work, we present the \gls{basmurf}, a \gls{sdn} routing scheme that takes into account the positioning uncertainty and beamforming design to find the most reliable routes in a \gls{fanet}. Our simulation results show that joint consideration of the beamforming and routing can provide a 5\% throughput improvement with respect to the state of the art.
\end{abstract}

\begin{IEEEkeywords}
Flying Ad-Hoc Networks, mmWave Communication, Beamforming, Position Uncertainty
\end{IEEEkeywords}

\glsresetall
\section{Introduction}
\glspl{uav} are used in a wide range of applications, from tracking and monitoring animals in remote areas\cite{gonzalez2016unmanned} to military applications\cite{ma2013simulation}. In order to effectively accomplish tasks over wider areas, it can be necessary to deploy multiple \glspl{uav}, which are expected to coordinate actions in an autonomous fashion or execute direct instructions from a control center, all of which are relayed through a \gls{fanet}.

In many scenarios, the \glspl{uav} need to exchange a relatively large amount of data with other members of the swarm and/or with the control station to support a given service. For example, distributed area monitoring/patrol applications may require the \glspl{uav} to stream high definition video or thermal camera recordings to the control station. Conversely, high bit rate data traffic demands wideband communication technologies (e.g., mmWave) that typically have limited coverage range, so that providing such services over wide areas may require multi-hop data connections, where the \glspl{uav} themselves can act as relays for other nodes in the network~\cite{ji2019performance}.

On the other hand, the \glspl{uav} and the control station also need to exchange light control traffic, which usually has strict latency and reliability constraints, but low bit rate requirements. This control channel can be used by the \glspl{uav} to send periodic tracking updates to the control center, which can use these messages to track the \gls{uav} positions \cite{wan2013smooth,biomo2014enhanced}. In these scenarios, the \glspl{uav} and the control center can use different technologies to carry information and signaling traffic, physically separating the data and control planes. The control traffic can be carried by low-rate long-range communication technologies such as LoRa \cite{farooq2021multi}, which can provide direct links between the \glspl{uav} and the control center. Following the \gls{sdn} paradigm, the control center, which has the most complete view of the state of the network, can determine and propagate routes in a centralized fashion through the same control channel~\cite{zhao2019software}.

Routing is a complex problem in \glspl{fanet}: even if the control center knows each \gls{uav}'s position at a given instant, the dynamic, three-dimensional nature of a swarm makes maintaining stable routes a difficult problem. Most routing protocols for \glspl{fanet} have been devised as an extension to the traditional \gls{manet} protocols such as \gls{aodv} and \gls{olsr}~\cite{katila2017routing,bahloul2018flocking,song2018mobility,xie2018enhanced,pari2018reliable}. %\FC{ANAY: can we add some references from the \gls{smurf} paper here? We're missing some literature review}

Moreover, the high bandwidth of mmWave technologies comes at the cost of a significantly higher path loss, reducing the effective range of communication. To mitigate this issue, mmWave systems often use beamforming techniques that can direct the signal towards the receiver, wasting less power on other directions and reducing interference. However, accurate beamforming requires an accurate knowledge of the transmitter's and receiver's positions, which is not always possible in a \gls{fanet}: the \glspl{uav} are moving, often at relatively high speed, and can only rely on imperfect sensors to measure their position. Additionally, sharing positioning information requires some signaling~\cite{mason2020combining}, which can be performed over long-range and low-bitrate technologies.

The use of beamforming with imperfect information introduces another challenge when making routing decisions, as \glspl{uav} at a shorter distance might have a higher probability of remaining in range, but also suffer more from beamforming errors: if the distance between two nodes is small, the effect of the positioning error on the beamforming angle is proportionally larger. In this work, we build on the \gls{smurf} scheme, which we presented in our previous work~\cite{deshpande2020smurf}, to build a scheme that can jointly consider routing and beamforming with uncertain position information. Our simulation results show that the proposed \gls{basmurf} scheme can increase the average throughput by 5\% over both \gls{smurf} and \gls{dbr}, an improvement that holds across different \gls{uav} network densities and antenna configurations.

The rest of this paper is divided as follows: Sec.~\ref{sec:system} presents the model of the swarm of \glspl{uav} as a network and the beamforming and channel model. Sec.~\ref{sec:routing} defines the routing design while considering beamforming and position uncertainty information. Sec.~\ref{sec:results} shows the simulation results for the \gls{basmurf} and its comparison with \gls{smurf} and \gls{dbr}. Sec.~\ref{sec:conclusion} explains the conclusions of the entire routing protocol.

\section{System Model}
\label{sec:system}
% Consider a scenario where $K$ \gls{uav}s have a set of existing links 

% Consider a scenario where $K$ \gls{uav}s are spread in the space and their connecti

We describe the connectivity in a \gls{fanet} as a time-varying graph $G = (K, E(t))$, where $K$ represents the set of \glspl{uav} in the network and $E(t)$ the set of active links at time $t$. Each drone moves independently from the rest and has its attitude characterized by a 5-tuple: the coordinates in space $\mathbf{x}_k(t)=\left(x_k(t), y_k(t), z_k(t)\right)$ and the yaw and pitch angles. So, the quality of the link between a pair of drones $(i,j)$ depends on their Euclidean distance \cite{deshpande2020smurf}: $$d_{ij}(t) = \lVert\mathbf{x}_i(t) - \mathbf{x}_j(t) \rVert_2.$$
Since \glspl{uav} fly high from the ground, the space between them is assumed to be free, i.e.,  without obstacles or reflections. We can then consider that there is predominantly \gls{los} communication \cite{al2020probability}. Thus, the pathloss between two \glspl{uav} is described by
\begin{equation}
    P_L(d_{ij})=\left(\frac{c}{4\pi f_0 d_{ij}}\right)^\gamma,
\end{equation}
where $f_0$ is the carrier frequency, $c$ is the speed of light, and $\gamma$ is the path loss exponent. As a result, if two \gls{uav}s are nearby, then there exists a link between them. Otherwise, communication becomes impractical due to low SNR. More formally, we can describe the set of active links by 
\begin{equation}
\label{eq:link:distance}
E(t) = \{e_{ij}(t): d_{ij} < D\},
\end{equation}
where the choice of $D$ often depends on the environment and the wireless technology used for communication \cite{noor2020review}.

% $$    d_{ij}(t)=||\mathbf{x}_i(t)-\mathbf{x}_j(t)||_2$$

% in the 3D space, described as $\mathbf{x}_k(t)=\left(x_k(t), y_k(t), z_k(t)\right)$.
% Each drone moves independently from each other and has its location, in the 3D space, described as $\mathbf{x}_k(t)=\left(x_k(t), y_k(t), z_k(t)\right)$. The link $e_{ij}(t)$ between a pair of drones $(i,j)$ can be described according to their euclidean distance \cite{deshpande2020smurf}. If two \gls{uav}s are nearby, then there exist a link between them. Otherwise, communication becomes impractical due to low SNR. More formally, we can describe the set of active links by 
% \begin{equation}

% \label{eq:link:distance}
% E(t) = \{e_{ij}(t): d_{ij} < R\}
% \end{equation}
% with $d_{ij} = ||\mathbf{p}_i(t) - \mathbf{p}_j(t) ||_2$.
% We emphasize that the choice of $R$ often depends on the environment and on the used communication technology \cite{noor2020review}.

% Let us first consider the case where the control station has knowledge over all positions $\mathbf{x}_k(t)$. In this case, the link $e_ij(t)$ between a pair of drones $(i,j)$ can be described according to their euclidean distance. If two \gls{uav}s are nearby, then there exist a link between them. Otherwise, communication becomes impractical. More formally, we can describe the set of active links by 
% $$
% E(t) = \{e_ij(t): ||\mathbf{p}_i(t) - \mathbf{p}_j(t) ||_2 < R\}.
% $$
% We emphasize that the choice of $R$ often depends on the environment and on the used communication technology \note{[REF]}.

Unfortunately, in more realistic scenarios, the real position of all drones is unknown, but the central controllers maintain estimators to predict the \glspl{uav}' positions based on current and previous location updates~\cite{mason2020combining}. Hence, the position estimate is defined as:
\begin{equation}
    \label{eq:position:noise}
    \hat{\mathbf{x}}_k(t) = \mathbf{x}_k(t) + \mathbf{n}_k(t),
\end{equation}
where the noise $\mathbf{n}_k(t)\sim\mathcal{N}(\mathbf{0}, \mathbf{\Sigma}_k)$ is associated to the $k$th user where $\mathbf{\Sigma}_k$ refers to the uncertainty covariance in the three coordinates. Hence, $\mathbf{\Sigma}_k$ is a symmetric matrix. Note that, even if $\mathbf{\Sigma}_k$ is a symmetric matrix, it is not an identity matrix, i.e. the error in estimate is not equal in all three coordinates. This is due to the fact that the error in estimate will be larger in the direction of movement for the \gls{uav} and smaller in other directions. In order to account for this uncertainty, one can also re-write (\ref{eq:link:distance}) in terms of the probability that the two endpoints are within a sphere of radius $D$. As a result, the link existence probability can be defined as
\begin{equation}
    \label{eq:link:probability}
    \hat{E}(t) = \left\{(i,j): i\neq j; P_{ij}(t) \right\}
\end{equation}
where the probability $P_{ij}(t)$ that the link between $i$ and $j$ is active at time $t$ is given by:
\begin{equation}\label{eq:link_existence}
P_{ij}(t) = \int_{\mathcal{B}_D(0)} \frac{e^{-\frac{1}{2} 
(\mathbf{x} - \mathbf{\Delta \hat{x}}_{ij}(t))^\mathrm{T}(\mathbf{\Sigma}_{ij}(t))^{-1}(\mathbf{x} - \mathbf{\Delta \hat{x}}_{ij}(t))}
}{2\pi|\mathbf{\Sigma}_{ij}(t)|^\frac{1}{2}} \,d\mathbf{x}.    
\end{equation}
Here, we have defined $\mathbf{\Delta \hat{x}}_{ij}(t) = \mathbf{\hat{x}}_i(t) - \mathbf{\hat{x}}_j(t)$, $\mathbf{\Sigma}_{ij}(t) = \mathbf{\Sigma}_{i}(t) + \mathbf{\Sigma}_j(t)$ and $\mathcal{B}_D(0)$ as the sphere with radius $D$ and center in the origin. 

\begin{figure}
    \centering
    \includegraphics[width=0.4\textwidth]{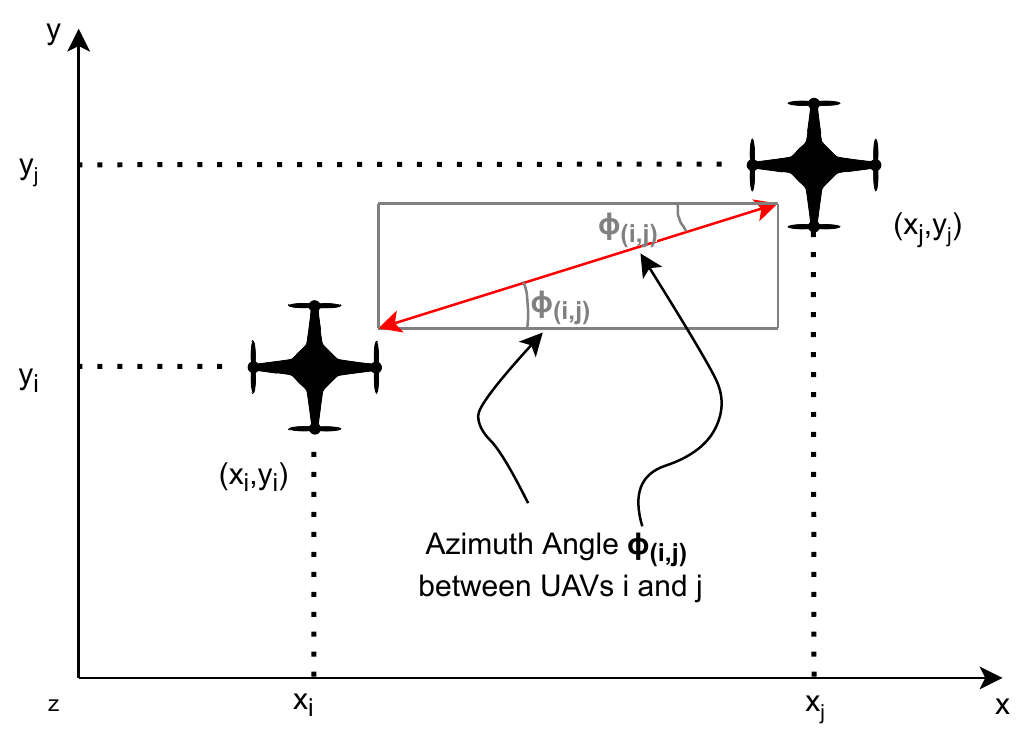}
    \caption{2-D representation of the angular separation between \glspl{uav} $i$ and $j$ for perfect knowledge of position and attitude.}
    \label{fig:drone_angle}
    % \vspace{-1\baselineskip}
\end{figure}

% However, we can 
To introduce beamforming into the system, we consider each \gls{uav} to be equipped with an \gls{upa} of dimension $ M = M_\mathrm{H} \times M_\mathrm{V}$, where antennas are spaced $d_\mathrm{H}$ (horizontally) and $d_\mathrm{V}$ (vertically) wavelengths from one another \cite{bjornson2017massive}. This allows each drone to communicate with other devices in the same altitude, as well as, with drones at different heights. As discussed above, due to the nature of the open environment, we assume that there is predominantly \gls{los} communication. In this scenario, the \glspl{uav} can also apply beamforming to improve the \gls{sinr} by increasing antenna gain. To determine the steering vector for beamforming, we need to determine the angular separation between the corresponding \glspl{uav}. \figurename~\ref{fig:drone_angle} shows the 2-D representation of the angular separation, i.e. azimuth angle, between the \glspl{uav} $i$ and $j$ for perfect knowledge of position and attitude. Since $\hat{\mathbf{x}}_i(t)$ and $\hat{\mathbf{x}}_{j}(t)$ are imperfect estimates, the angle between the estimated position $\hat{\mathbf{x}}_{j}(t)$ of \gls{uav} $j$ with respect to $\hat{\mathbf{x}}_i(t)$ of \gls{uav} $i$ is itself a random variable. In the following, we omit the time index $t$ for simplicity. 
Hence, the steering vector is defined based on the estimated azimuth $\Delta\hat{\phi}_{ij}$ and elevation $\Delta\hat{\theta}_{ij}$ angles between the $j$th and $i$th \glspl{uav}, respectively, described by,
\begin{align}
   \Delta\hat{\phi}_{ij}&=\pi\mathbb{1}(\hat{x}_j-\hat{x}_i)+\arctan\left(\frac{\hat{y}_j-\hat{y}_i}{\hat{x}_j-\hat{x}_i}\right)-\hat{\phi}_i;\\
   \Delta\hat{\theta}_{ij}&=\arctan\left(\frac{\hat{z}_j-\hat{z}_i}{\sqrt{(\hat{x}_j-\hat{x}_i)^2+(\hat{y}_j-\hat{y}_i)^2}}\right)-\hat{\theta}_i,
\end{align}
where $\mathbb{1}(x)$ is the step function, equal to $1$ if $x>0$ and $0$ otherwise and $\hat{\phi}_i$ and $\hat{\theta}_i$ refers to the estimated yaw and pitch angles of the \gls{uav}. %Similarly, we can define the true angles $\Delta\phi_{ij}$ and $\Delta\theta_{ij}$ by using the true position and attitude vectors of the two \glspl{uav}. %Hence, the azimuth deviation $\psi_{ij}$ and elevation deviation $\omega_{ij}$ from the real angle are:
%\begin{align}
%        \psi_{ij} = \Delta\hat{\phi}_{ij}-\Delta\phi_{ij};\\
%    \omega_{ij} = \Delta\hat{\theta}_{ij}-\Delta\theta_{ij}.
%\end{align}
We can then compute the beamforming gain due to uncertainty in position information as a function of $\Delta\hat{\phi}_{ij}$ and $\Delta\hat{\theta}_{ij}$:
\begin{align}\label{eq:omega}
    g(\Delta\hat{\phi}_{ij}, \Delta\hat{\theta}_{ij}) = \frac{1}{M_\mathrm{H} M_\mathrm{V}}\frac{1}{\mathbf{\Delta x}_{ij}}\mathrm{h}_{Rx}\mathrm{h}_{Tx},
\end{align}
where,
\begin{align}
    \mathrm{h}_{Tx} &= |\mathbf{a}(\Delta\hat{\phi}_{ij}, \Delta\hat{\theta}_{ij})^\mathrm{H}\mathbf{w}_{Tx}|,\\
    \mathrm{h}_{Rx} &= |\mathbf{a}(\pi+\Delta\hat{\phi}_{ij}, \pi+\Delta\hat{\theta}_{ij})^\mathrm{H}\mathbf{w}_{Rx}|,
\end{align}
where $\mathbf{w}_{Tx}/ \mathbf{w}_{Rx} \in \mathbb{C}^{M\times 1}$ denotes the beamforming vector used for transmission/reception and
\begin{equation}
    \label{eq:beamsteer}
    \mathbf{a}(\phi, \theta) = \left[e^{j\mathbf{\kappa}^\mathrm{T}(\phi, \theta)\mathbf{u}_1}, \dots, e^{j\mathbf{\kappa}^\mathrm{T}(\phi, \theta)\mathbf{u}_M} \right]^\mathrm{T},
\end{equation}
% \left[e^{i(l)d_H\cos(\theta)\sin(\phi)},~e^{j(l)d_V\sin(\theta)}\right]^\mathrm{T},
is the steering vector matrix associated to the azimuth ($\phi$) and elevation ($\theta$) angles.  Also, note that, the azimuth for the $j^{th}$ \gls{uav} with respect to $i^{th}$ \gls{uav} is opposite of the angles of $i^{th}$ \gls{uav} with respect to $j^{th}$ \gls{uav}, i.e., these angles are supplementary due to the central controller devising the directions the \glspl{uav} have to orient themselves for transmission/reception. The same analogy is valid for the elevation angles between these UAVs. For simplicity, we have also defined $\mathbf{\kappa}(\alpha, \theta)$ as the wave vector for a planar wave impinging with angles $\phi$ and $\theta$, $\lambda$ is the wavelength and $\mathbf{u}_{m}$ is the 3D spatial location of the $m$th element of the antenna array. Specifically, for \gls{upa}, $\mathbf{u}_m = \left[0,~i(m)d_H\lambda,~ j(m)d_V\lambda \right]^\mathrm{T}$  where we also consider the auxiliary functions 
$i(l) = \mathrm{mod}(l - 1, M_\mathrm{H})$ and $j(l) = \lfloor(l - 1)/M_\mathrm{V}\rfloor$.

% Also, note that, the azimuth and elevation angles for the $j^{th}$ \gls{uav} with respect to $i^{th}$ \gls{uav} is opposite of the angles of $i^{th}$ \gls{uav} with respect to $j^{th}$ \gls{uav}.
% For simplicity, we have also defined $\mathbf{\kappa}(\alpha, \theta)$ as the wave vector for a planar wave impinging with angles $\alpha$ and $\theta$, $\lambda$ is the wavelength and $\mathbf{u}_{m}$ the 3D spatial location of the $m$th element of the antenna array.

% $\mathbf{a}(\phi,\theta) \in \mathbb{C}^{M \times 2}$ denotes the steering vector associated to the azimuth and elevation angles, with $M$ columns
% \begin{equation}
%     \label{eq:beamsteer}
%     a_l(\phi, \theta) = 2\pi\left[e^{i(l)d_H\cos(\theta)\sin(\phi)},~e^{j(l)d_V\sin(\theta)}\right]^\mathrm{T},
% \end{equation}
% where $d_H=d_V=\frac{1}{2}$. 

% Also, note that, the azimuth and elevation angles for the $j^{th}$ \gls{uav} with respect to $i^{th}$ \gls{uav} is opposite of the angles of $i^{th}$ \gls{uav} with respect to $j^{th}$ \gls{uav}.For simplicity's sake, we  have also defined the auxiliary functions 

% where $d_H$ and $d_V$ are $\frac{1}{2}$. For simplicity's sake, we  have also defined the auxiliary functions 
% $
% i(l) = \mathrm{mod}(l - 1, M_\mathrm{H})$ and $j(l) = \lfloor(l - 1)/M_\mathrm{V}\rfloor
% $. 
%Similarly, the beamforming gain can be defined for true angles $\Delta\phi_{ij}$ and $\Delta\theta_{ij}$ using (\ref{eq:omega}).
We can then compute the expected received power $P^{(r)}_{ij}$ over link $(i,j)$ as:
\begin{equation}\label{eq:power}
    P^{(r)}_{ij} = \frac{P_{\text{Tx}}|g(\Delta\hat{\phi}_{ij}, \Delta\hat{\theta}_{ij})|^2}{P_L(d_{ij})}.
\end{equation}

The received power only depends on the transmission power $P_{\text{Tx}}$, the beamforming gain for the transmission and reception, and the distance $d_{ij}$. Note that, the central controller devises the transmit and receive beamforming vector $\mathbf{w}_{Rx}$ based on the estimated positions of the \glspl{uav} and thereby determines the expected received power at the receiver. By using the Shannon rate formula, we can then get the expected capacity $C_{ij}(t)$ under interference
\begin{equation}\label{eq:capacity}
    C_{ij}(t)=B\log_2\left(1+\frac{P^{(r)}_{ij}(t)}{\sum\limits_{\ell \neq i} P^{(r)}_{\ell j}(t) + N_0 B}\right),
\end{equation}
where $N_0$ is the noise power spectral density and $B$ the bandwidth of the channel. Naturally, the distribution of this capacity is extremely complex, as it is a highly nonlinear function of the estimated positions and attitudes in the swarm. As such, it is extremely hard to estimate directly, but we can use a Monte Carlo sampling method to draw the real state of the swarm from the belief distribution, which can approximate the real distribution when given enough samples.

\subsection{Analog Beamforming Design}

The design of the beamforming vectors $\mathbf{w}_{Tx}$ and $\mathbf{w}_{Rx}$ directly influences the behaviour of the power response $P^{(r)}_{ij}$. For a fixed transmitter to simultaneously communicate with multiple receivers, it is essential to design a beamforming that directs the emitted signal towards these receivers while keeping the interference from other receivers as low as possible. For instance, when the \gls{csi} of all receivers is perfectly known at the transmitter, one can design $\mathbf{w}_{Tx}$ and $\mathbf{w}_{Rx}$ as a zero-forcing beamformers \cite{bjornson2017massive}. %The effective design of $\mathbf{W}_{ij}$ and usage of the spatial domain is an active line of research in massive MIMO  \note{[RSMAC, JSDM, mMIMO beamforming]}. % is this last line necessary?
% However, there are
This type of beamforming design is primarily applicable for traditional fixed-transmitter communication systems. Our scenario, however, differs in at least three aspects from this traditional scheme.  Firstly, each \gls{uav} is only interested in communicating with a single receiver at a particular time based on the route devised by the routing protocol. Secondly, the control station does not have the current \gls{csi} for each of the \glspl{uav}. In fact, keeping track of all \glspl{csi} among the different \glspl{uav} in this scenario is a hard task due to high mobility of the \glspl{uav}. Finally and the most important characteristic of the scenario is the mobility of the transmitter. In our scenario, \glspl{uav} are mobile transmitters which can easily be rotated towards a desired direction. Thus, in our devised scenario, alignment happens by rotating pairs of \glspl{uav} towards each other. 

Moreover, due to the characteristics mentioned above, designing $\mathbf{w}_{Tx}$ and $\mathbf{w}_{Rx}$ for our communication scheme boils down to choosing a narrow or wide beam. Notice that, in the former, a narrow beam potentially reduces the amount of interference leaked towards other \glspl{uav}, but requires knowledge over the true position $\mathbf{x}_k$. In contrast, designing a wide beam can compensate for the uncertainty over the desired location while potentially increasing interference among different \glspl{uav}. We simulate this behaviour by turning on/off the last rows or columns of the \gls{upa} of each \gls{uav}. To simulate this idea, let us consider the logical matrix $\mathbf{W}$ with entries $\{0,1\}$ and dimension ${M_H\times M_V}$. We can design the beamforming vector $\mathbf{w}$ by stacking the columns of matrix $\mathbf{W}$, i.e., % is mapped to a beamforming vector by stacking its columns, i.e., %$\mathbf{w} = [\mathbf{W}_1, \mathbf{W}_2, \dots, \mathbf{W}_{M_V}]^\text{T}$.
\begin{equation}
\mathbf{w} = \begin{bmatrix}
    \mathbf{W}_1      \\
   \mathbf{W}_2 \\
   \vdots  \\
   \mathbf{W}_V
\end{bmatrix}
\end{equation}
where $\mathbf{W}_{k}$ denotes the $k$th column of the logical matrix.
%$$
%\mathbf{w} = \begin{bmatrix}
%    \mathbf{W}_1      \\
%   \mathbf{W}_2 \\
%   \vdots  \\
%   \mathbf{W}_V
%\end{bmatrix}.
%$$
Then choosing a wide beam translates into setting the last rows/columns of the matrix $\mathbf{W}$ to zero and the remaining entries to one. For instance, to have a omnidirectional transmission, it is sufficient to have a single active antenna~\cite{qiao2016broadbeam}. Opposite to that, setting all entries of $\mathbf{W}$ to one results in the narrow beam pattern. %, in our scenario this means $\mathbf{W}_{1,1} = 1$ and $\mathbf{W}_{m,l} = 0, \forall m,l\neq 1$
Using this idea we derive wider and narrower beams during our simulations. Moreover, we assume every transmitter to use the same beamformer over the entire route.

% Let us first consider the decomposed spatial filter $\mathbf{w} = \mathbf{w}_{H} \cdot \mathbf{w}_{V}$ as the dot product of its horizontal ($\mathbf{w}_{H}$) and vertical ($\mathbf{w}_{V}$) contributions. Moreover, let these contributions have binary entries. Then choosing a wider or narrower beam translates into choosing the entries of $\mathbf{w}_{H}, \mathbf{w}_{V}$ to be a combination of zeros and ones.
% For instance, to have a omnidirectional transmission, it is sufficient to have a single  active antenna~\cite{qiao2016broadbeam}. Using this idea we derive wider and narrower beams during our simulations. Moreover, we assume every transmitter to use the same beamformer $\mathbf{w}$ over the entire route.

% \note{TODO:} In terms of the beamformer $\mathbf{w}$, this is represented by defining its entries as binary quantities $[\mathbf{w}]_{l,m} \in \{0,1\}$. For instance, to have a omnidirectional transmission, it is sufficient to have a single active antenna \cite{qiao2016broadbeam}. For example, $l=\frac{M_\mathrm{H}}{2}, m=1$ represents a horizontal central antenna which performs an omnidirectional transmission. Moreover, we assume every transmitter to use the same beamformer $\mathbf{W}$ over the entire route. 

Finally, we also assume that there is no interference between the routes, that have common \glspl{uav} as their intermediate relays, by defining a simplified \gls{tdma} mechanism which divides the total time of communication equally among all different routes for the common \gls{uav}. This \gls{tdma} mechanism is defined by the control station, which knows all the concurrent routes in the network at a particular time.

\section{Position Uncertainty Based Beamformed Routing}\label{sec:routing}

Using the beamforming vectors defined in the previous section, we define the \gls{basmurf} protocol. In this protocol, the edge weight $\hat{e}_{ij}\in \hat{E}(t)$ represents the average expected capacity of the link between \glspl{uav} $i$ and $j$, denoted by $C_{ij}(t)$. Hence, we design this routing problem as a standard maximum capacity route problem~\cite{hu1961letter} over the graph $\hat{G}=(K,\hat{E}(t))$, which can be solved by determining the maximum spanning tree using the link capacities, weighted by the source load, as a weighting metric for the graph edges. So, for a given beamformer, we apply a maximum spanning tree on the network graph $G$ to determine the route with maximum achievable capacity.

%Additionally, using the link existence probability from (\ref{eq:link_existence}), we can deploy a similar procedure to the \gls{smurf} routing scheme~\cite{deshpande2020smurf} %\FC{Define acronym} to compute the routes:
%\begin{itemize}
%    \item The central controller calculates the link existence probability between all the \glspl{uav} in the network using (\ref{eq:link_existence}).
%    \item It then determines the conditional link existence probability between three consecutive \glspl{uav}. For example, for the consecutive \glspl{uav} $i \rightarrow j \rightarrow k$, 
    %For example, for \glspl{uav} $i,j$ and $k$, 
%    it determines the probability of link existence between $j$ and $k$ given that the link exists between $i$ and $j$.
%    \item To determine the route existence probability, the controller has to calculate the conditional probability of link existence given that all the previous links exists. But in realistic scenarios, the conditional probability only depends on the positions of \glspl{uav} in consideration. Thereby, the route existence probability is determined by,
%$$
%Pr(\mathbf{e}) \simeq Pr(e_{12})Pr(e_{23}|Pr_{12})\ldots Pr(e_{n-1,n}|e_{n-2,n-1}).\label{eq:route_p}
%$$
%    \item Once, the route existence probability is calculated for all routes for a particular source-destination pair, the route with highest probability is chosen as the primary route. The protocol also determines backup routes for the primary route but those have been excluded for this work.
%\end{itemize}

% brautiful

Hence, using (\ref{eq:capacity}), we devise the routes that maximize the minimum expected capacity. We assume that each \gls{uav} $i$ is occupied in transmitting or receiving cross-traffic for a fraction $\rho_i$ of the time i.e. $\rho_i = [0,1]$, which must be subtracted to determine the available capacity of the link for \gls{uav} $i$. As transmission is not full-duplex, we also assume that each node can spend at most half the time transmitting. So, the expected capacity of each link can be approximated by Monte Carlo sampling, and we can build a graph, knowing that the capacity of a route $\mathbf{r}$ is given by:
\begin{equation}
    C(\mathbf{r})=\min_{(i,j)\in \mathbf{r}} \frac{(1-\rho_i)C_{i,j}}{2}.
\end{equation}
%RE-Written: So, to determine route with maximum achievable capacity for a route based on a given beamforming matrix, we apply a maximum spanning tree on the network graph $G$.

The entire routing algorithm is devised at the central controller which tracks the positions of all the \glspl{uav} \cite{mason2020combining}. This, however, limits the beamforming adaptability over a single route, i.e., all the \glspl{uav} in the route follow the same beamforming pattern. In this work, the central controller devises the beamforming patterns for all the \glspl{uav} in the entire route based on the estimated position of the receiver \gls{uav} with respect to transmitter \gls{uav}, i.e., the central controllers designs the beamsteering vector shown in (\ref{eq:beamsteer}). In principle, it might seem sub-optimal to fix the beamforming pattern and only design the beamsteering vector (i.e., deploy analog beamforming), but to devise beamformed routes, the central controller needs to have a perfect instantaneous knowledge of the channel (i.e., deploy digital beamforming), which is difficult in the current scenario.
%We can then pose two routing problems:
%\begin{enumerate}
%    \item The \emph{maximum expected capacity} problem is to find the route that maximizes the expected capacity given the positioning information and uncertainty;
%    \item The \emph{maximum $\varepsilon$-guaranteed capacity} problem is to find the route that maximizes the capacity $C_{\varepsilon}$ such that $P(C(\mathbf{r})\geq C_{\varepsilon})\geq1-\varepsilon$.
%\end{enumerate}
%Both of these problems are standard maximum capacity route problems, which can be solved by finding the maximum spanning tree using the link capacities, weighted by the source load, as a weighting metric for the graph edges.

%\FC{We should say something here on WHO exactly does the routing: is it a central controller, with full info on all UAVs? Is it the source? Is it each UAV individually? Also, how do we compute W together with the route? Are we adapting beamforming or is it just a beamforming-aware protocol?}

\section{Results} \label{sec:results}
\begin{table}
\centering
\caption{Parameters of the Simulations}
\begin{tabular}{|c|c|}
\hline
Simulation Parameter         & Simulation Value            \\ \hline
Map Size [m]                & $200\times 200\times 10  $         \\ \hline
Density of UAVs [UAVs/km$^3$]           & $\{25000-75000\}$         \\
\hline
UAV Position Model           & Unscented Kalman Filter     \\ \hline
Maximum Transmission Distance [m]       & 100                        \\ \hline
MIMO Antenna                & Uniform Planar Array (UPA) \\ \hline
Antenna Configurations for the UAVs           & $\{1,4,8,16,32,64\}$         \\
\hline
Transmission Power [W]          & 1                       \\ \hline
Number of Simulated Networks & 240                         \\ \hline
Bandwidth (MHz) & 100                         \\ \hline
\end{tabular}
\label{tab:simulation}
\end{table}
In order to numerically evaluate the effects of position uncertainty and beamforming in routing protocols, we deploy a Monte Carlo simulation in MATLAB where the protocols are evaluated over randomly generated networks. The simulation parameters for the system are described in Tab. \ref{tab:simulation}. The protocols evaluated in the Monte Carlo simulation are:
\begin{enumerate}
    \item \gls{dbr}: A purely distance-based protocol, which does not consider positioning uncertainty;
    \item \gls{smurf}: our scheme from~\cite{deshpande2020smurf}, limited to a single path, which considers positioning uncertainty but does not include beamforming in the probability calculation;
    \item \gls{basmurf}: the proposed scheme, which takes into account the position uncertainty and beamforming.
\end{enumerate}
% 
% \begin{enumerate}
%     \item \gls{smurf}-T: \gls{smurf} protocol~\cite{deshpande2020smurf} limited to primary path considering only tracked position;
%     \item \gls{dbr}-T: Purely distance based protocol with only tracked position information;
%     \item \gls{basmurf}-T: \gls{basmurf} protocol with only tracked position information and fixed beamforming design;
%     \item \gls{smurf}: \gls{smurf} protocol limited to primary path with true position information;
%     \item \gls{dbr}: Purely distance based protocol with true position information;
%     \item \gls{basmurf}: \gls{basmurf} protocol with true position information.
% \end{enumerate}
These protocols are evaluated for both the ideal and tracked position. We indicate the former by attaching ``-I'' to the end of the respective protocol name, and the latter by attaching ``-T'. Moreover, the tracked position information is obtained from the output of the Kalman filter at the central controller. Hence, the performance of the protocols evaluated in this scenario is the performance achieved based on the available tracked information of the \glspl{uav}. By taking into account the true position of the \glspl{uav}, we can determine the ideal performance of the protocol, which represents an upper bound to the practically achievable performance. Additionally, each \gls{uav} is equipped with the same \gls{upa} antenna configuration and is able to communicate with other drones using the beamforming design defined in the previous section.

% The first three protocols considered for evaluation take into account the tracked position information which is obtained from the output of the Kalman filter at the central controller. Hence, the performance obtained for the first three protocols is the estimated performance calculated by the central controller based on the available tracked information of the \glspl{uav}. The next three protocols take into account the true position of the \glspl{uav} and determine the ideal performance of the protocol considering perfect position information. Additionally, each \gls{uav} is equipped with the same UPA antenna configuration and is able to communicate with each other using the beamforming design as defined in the previous section. 
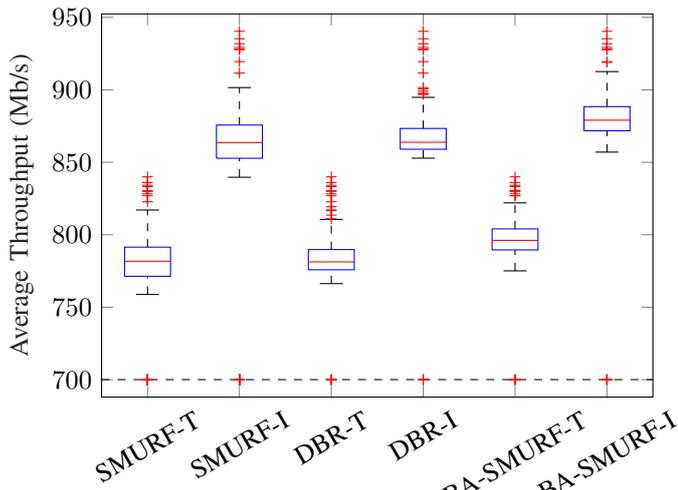
\begin{figure}
    \centering
\begin{tikzpicture}

\begin{axis}[%
width=0.4\textwidth,
height=2in,
at={(0in,0in)},
scale only axis,
xmin=0.5,
xmax=6.5,
xtick={1,2,3,4,5,6},
xticklabels={{SMURF-T},{SMURF-I},{DBR-T},{DBR-I},{BA-SMURF-T},{BA-SMURF-I}},
xticklabel style={rotate=30},
ymin=687.980442075993,
ymax=952.410716404138,
%axis background/.style={fill=white}
ylabel style={font=\color{white!15!black}},
ylabel={Average Throughput (Mb/s)},
]
\addplot [color=black, dashed, forget plot]
  table[row sep=crcr]{%
0.5	700\\
6.5	700\\
};
\addplot [color=black, dashed, forget plot]
  table[row sep=crcr]{%
1	791.439404848961\\
1	817.076125443061\\
};
\addplot [color=black, dashed, forget plot]
  table[row sep=crcr]{%
2	875.702408584081\\
2	901.556070386678\\
};
\addplot [color=black, dashed, forget plot]
  table[row sep=crcr]{%
3	789.772577961501\\
3	810.589473536863\\
};
\addplot [color=black, dashed, forget plot]
  table[row sep=crcr]{%
4	873.352341545827\\
4	894.83792460596\\
};
\addplot [color=black, dashed, forget plot]
  table[row sep=crcr]{%
5	804.028074767099\\
5	821.995716379999\\
};
\addplot [color=black, dashed, forget plot]
  table[row sep=crcr]{%
6	888.427204122716\\
6	912.621094034347\\
};
\addplot [color=black, dashed, forget plot]
  table[row sep=crcr]{%
1	758.796449283414\\
1	771.344991389058\\
};
\addplot [color=black, dashed, forget plot]
  table[row sep=crcr]{%
2	839.687828834521\\
2	852.79556817191\\
};
\addplot [color=black, dashed, forget plot]
  table[row sep=crcr]{%
3	766.266822718645\\
3	775.838322602133\\
};
\addplot [color=black, dashed, forget plot]
  table[row sep=crcr]{%
4	852.858531214555\\
4	859.015426495997\\
};
\addplot [color=black, dashed, forget plot]
  table[row sep=crcr]{%
5	775.085321036691\\
5	789.527802265623\\
};
\addplot [color=black, dashed, forget plot]
  table[row sep=crcr]{%
6	857.045138285644\\
6	871.762354359595\\
};
\addplot [color=black, forget plot]
  table[row sep=crcr]{%
0.875	817.076125443061\\
1.125	817.076125443061\\
};
\addplot [color=black, forget plot]
  table[row sep=crcr]{%
1.875	901.556070386678\\
2.125	901.556070386678\\
};
\addplot [color=black, forget plot]
  table[row sep=crcr]{%
2.875	810.589473536863\\
3.125	810.589473536863\\
};
\addplot [color=black, forget plot]
  table[row sep=crcr]{%
3.875	894.83792460596\\
4.125	894.83792460596\\
};
\addplot [color=black, forget plot]
  table[row sep=crcr]{%
4.875	821.995716379999\\
5.125	821.995716379999\\
};
\addplot [color=black, forget plot]
  table[row sep=crcr]{%
5.875	912.621094034347\\
6.125	912.621094034347\\
};
\addplot [color=black, forget plot]
  table[row sep=crcr]{%
0.875	758.796449283414\\
1.125	758.796449283414\\
};
\addplot [color=black, forget plot]
  table[row sep=crcr]{%
1.875	839.687828834521\\
2.125	839.687828834521\\
};
\addplot [color=black, forget plot]
  table[row sep=crcr]{%
2.875	766.266822718645\\
3.125	766.266822718645\\
};
\addplot [color=black, forget plot]
  table[row sep=crcr]{%
3.875	852.858531214555\\
4.125	852.858531214555\\
};
\addplot [color=black, forget plot]
  table[row sep=crcr]{%
4.875	775.085321036691\\
5.125	775.085321036691\\
};
\addplot [color=black, forget plot]
  table[row sep=crcr]{%
5.875	857.045138285644\\
6.125	857.045138285644\\
};
\addplot [color=blue, forget plot]
  table[row sep=crcr]{%
0.75	771.344991389058\\
0.75	791.439404848961\\
1.25	791.439404848961\\
1.25	771.344991389058\\
0.75	771.344991389058\\
};
\addplot [color=blue, forget plot]
  table[row sep=crcr]{%
1.75	852.79556817191\\
1.75	875.702408584081\\
2.25	875.702408584081\\
2.25	852.79556817191\\
1.75	852.79556817191\\
};
\addplot [color=blue, forget plot]
  table[row sep=crcr]{%
2.75	775.838322602133\\
2.75	789.772577961501\\
3.25	789.772577961501\\
3.25	775.838322602133\\
2.75	775.838322602133\\
};
\addplot [color=blue, forget plot]
  table[row sep=crcr]{%
3.75	859.015426495997\\
3.75	873.352341545827\\
4.25	873.352341545827\\
4.25	859.015426495997\\
3.75	859.015426495997\\
};
\addplot [color=blue, forget plot]
  table[row sep=crcr]{%
4.75	789.527802265623\\
4.75	804.028074767099\\
5.25	804.028074767099\\
5.25	789.527802265623\\
4.75	789.527802265623\\
};
\addplot [color=blue, forget plot]
  table[row sep=crcr]{%
5.75	871.762354359595\\
5.75	888.427204122716\\
6.25	888.427204122716\\
6.25	871.762354359595\\
5.75	871.762354359595\\
};
\addplot [color=red, forget plot]
  table[row sep=crcr]{%
0.75	781.726054236127\\
1.25	781.726054236127\\
};
\addplot [color=red, forget plot]
  table[row sep=crcr]{%
1.75	863.629142189302\\
2.25	863.629142189302\\
};
\addplot [color=red, forget plot]
  table[row sep=crcr]{%
2.75	781.269874092945\\
3.25	781.269874092945\\
};
\addplot [color=red, forget plot]
  table[row sep=crcr]{%
3.75	863.810999567334\\
4.25	863.810999567334\\
};
\addplot [color=red, forget plot]
  table[row sep=crcr]{%
4.75	796.103071034134\\
5.25	796.103071034134\\
};
\addplot [color=red, forget plot]
  table[row sep=crcr]{%
5.75	879.05898946951\\
6.25	879.05898946951\\
};
\addplot [color=black, only marks, mark=+, mark options={solid, draw=red}, forget plot]
  table[row sep=crcr]{%
1	700\\
1	700\\
1	700\\
1	700\\
1	700\\
1	700\\
1	700\\
1	700\\
1	700\\
1	700\\
1	700\\
1	700\\
1	700\\
1	700\\
1	700\\
1	700\\
1	700\\
1	700\\
1	700\\
1	700\\
1	700\\
1	700\\
1	700\\
1	700\\
1	700\\
1	700\\
1	700\\
1	700\\
1	700\\
1	700\\
1	700\\
1	700\\
1	700\\
1	700\\
1	700\\
1	700\\
1	700\\
1	822.751882283627\\
1	826.874189829986\\
1	828.234460223951\\
1	829.933094711394\\
1	830.420101154119\\
1	833.227166172048\\
1	834.086314969627\\
1	836.057745818169\\
1	839.979881553965\\
};
\addplot [color=black, only marks, mark=+, mark options={solid, draw=red}, forget plot]
  table[row sep=crcr]{%
2	700\\
2	700\\
2	700\\
2	700\\
2	700\\
2	700\\
2	700\\
2	700\\
2	700\\
2	700\\
2	700\\
2	700\\
2	700\\
2	700\\
2	700\\
2	700\\
2	700\\
2	700\\
2	700\\
2	700\\
2	700\\
2	700\\
2	700\\
2	700\\
2	700\\
2	700\\
2	700\\
2	700\\
2	700\\
2	700\\
2	700\\
2	700\\
2	700\\
2	700\\
2	700\\
2	700\\
2	700\\
2	911.584648301343\\
2	919.44168010177\\
2	927.765063857398\\
2	927.970140641721\\
2	929.242107238663\\
2	931.805307489504\\
2	935.277150594486\\
2	940.391158480132\\
};
\addplot [color=black, only marks, mark=+, mark options={solid, draw=red}, forget plot]
  table[row sep=crcr]{%
3	700\\
3	700\\
3	700\\
3	700\\
3	811.032802562852\\
3	813.531780713799\\
3	816.758335685663\\
3	817.076125443061\\
3	819.373181808825\\
3	822.751882283627\\
3	826.874189829986\\
3	828.234460223951\\
3	829.933094711394\\
3	830.420101154119\\
3	833.227166172048\\
3	834.086314969627\\
3	836.057745818169\\
3	839.979881553965\\
};
\addplot [color=black, only marks, mark=+, mark options={solid, draw=red}, forget plot]
  table[row sep=crcr]{%
4	700\\
4	700\\
4	700\\
4	700\\
4	896.854752621177\\
4	897.179376899083\\
4	897.601198966069\\
4	899.135276833921\\
4	900.609510797514\\
4	901.556070386678\\
4	911.584648301343\\
4	919.44168010177\\
4	927.765063857398\\
4	927.970140641721\\
4	929.242107238663\\
4	931.805307489504\\
4	935.277150594486\\
4	940.391158480132\\
};
\addplot [color=black, only marks, mark=+, mark options={solid, draw=red}, forget plot]
  table[row sep=crcr]{%
5	700\\
5	700\\
5	700\\
5	700\\
5	826.874189829986\\
5	827.006733562006\\
5	828.234460223951\\
5	829.933094711394\\
5	830.420101154119\\
5	830.593827835534\\
5	833.227166172048\\
5	834.086314969627\\
5	836.057745818169\\
5	839.979881553965\\
};
\addplot [color=black, only marks, mark=+, mark options={solid, draw=red}, forget plot]
  table[row sep=crcr]{%
6	700\\
6	700\\
6	700\\
6	700\\
6	919.001844628814\\
6	919.44168010177\\
6	927.765063857398\\
6	927.970140641721\\
6	929.242107238663\\
6	931.805307489504\\
6	935.277150594486\\
6	940.391158480132\\
};
\end{axis}
\end{tikzpicture}%
    \vspace{-2\baselineskip}
    \caption{Boxplot of the average throughput obtained for all the protocols in different experiments with a network density of $50000$~UAVs/km$^3$ and $M=16$ antenna elements.}
    \label{fig:tracked_throughput}
    \vspace{-1.\baselineskip}
\end{figure}
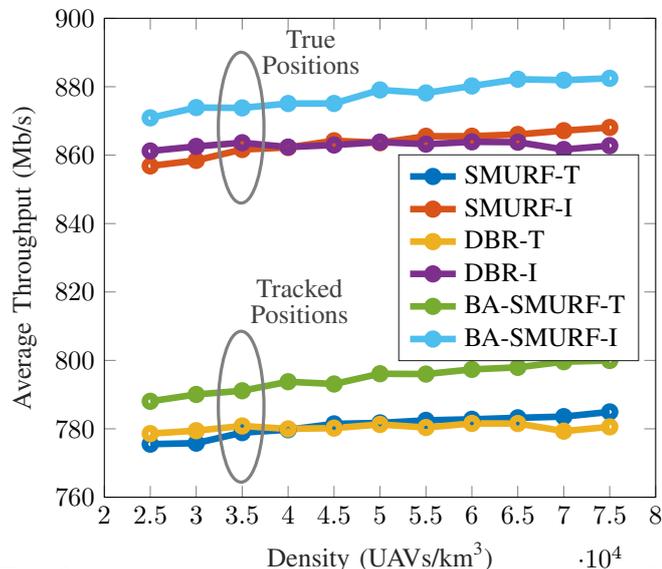
\begin{figure}
    \centering
\definecolor{mycolor1}{rgb}{0.00000,0.44700,0.74100}%
\definecolor{mycolor2}{rgb}{0.85000,0.32500,0.09800}%
\definecolor{mycolor3}{rgb}{0.92900,0.69400,0.12500}%
\definecolor{mycolor4}{rgb}{0.49400,0.18400,0.55600}%
\definecolor{mycolor5}{rgb}{0.46600,0.67400,0.18800}%
\definecolor{mycolor6}{rgb}{0.30100,0.74500,0.93300}%
\begin{tikzpicture}

\begin{axis}[%
width=0.4\textwidth,
height=2.5in,
at={(0in,0in)},
scale only axis,
xmin=20000,
xmax=80000,
xtick={20000, 25000, 30000, 35000, 40000, 45000, 50000, 55000, 60000, 65000, 70000, 75000, 80000},
xlabel style={font=\color{white!15!black}},
xlabel={$\text{Density (UAVs/km}^\text{3}\text{)}$},
ymin=760,
ymax=900,
ylabel style={font=\color{white!15!black}},
ylabel={Average Throughput (Mb/s)},
axis background/.style={fill=white},
legend style={at={(0.97,0.5)}, anchor=east, legend cell align=left, align=left, draw=white!15!black}
]
\addplot [color=mycolor1, line width=2.5pt, mark=o, mark options={solid, mycolor1}]
  table[row sep=crcr]{%
25000	775.518333274581\\
30000	775.805260314138\\
35000	778.883792770847\\
40000	779.732752064042\\
45000	781.436487628206\\
50000	781.726054236127\\
55000	782.470732972686\\
60000	782.777777590559\\
65000	783.215195623166\\
70000	783.563097383206\\
75000	784.93901148736\\
};
\addlegendentry{SMURF-T}

\addplot [color=mycolor2, line width=2.5pt, mark=o, mark options={solid, mycolor2}]
  table[row sep=crcr]{%
25000	856.841016587788\\
30000	858.461912637254\\
35000	861.642088394091\\
40000	862.241260819104\\
45000	864.201600665887\\
50000	863.629142189302\\
55000	865.516814297897\\
60000	865.536053797328\\
65000	866.046490815845\\
70000	867.146882073891\\
75000	868.090800766229\\
};
\addlegendentry{SMURF-I}

\addplot [color=mycolor3, line width=2.5pt, mark=o, mark options={solid, mycolor3}]
  table[row sep=crcr]{%
25000	778.59675724617\\
30000	779.458391746857\\
35000	780.831442557426\\
40000	779.976346052199\\
45000	780.203554967953\\
50000	781.269874092945\\
55000	780.374455480415\\
60000	781.569796355874\\
65000	781.540553834655\\
70000	779.317232777467\\
75000	780.575338313849\\
};
\addlegendentry{DBR-T}

\addplot [color=mycolor4, line width=2.5pt, mark=o, mark options={solid, mycolor4}]
  table[row sep=crcr]{%
25000	861.209870849371\\
30000	862.54473831093\\
35000	863.629059078908\\
40000	862.353319438787\\
45000	862.956349401108\\
50000	863.810999567334\\
55000	863.210895559612\\
60000	863.885888211351\\
65000	863.785802192281\\
70000	861.652003874067\\
75000	862.77260390747\\
};
\addlegendentry{DBR-I}

\addplot [color=mycolor5, line width=2.5pt, mark=o, mark options={solid, mycolor5}]
  table[row sep=crcr]{%
25000	788.042252876723\\
30000	790.040716083732\\
35000	791.12406713751\\
40000	793.762058020733\\
45000	793.102300235726\\
50000	796.103071034134\\
55000	795.999604163986\\
60000	797.369112866291\\
65000	797.961637585654\\
70000	799.612478728166\\
75000	799.959995118811\\
};
\addlegendentry{BA-SMURF-T}

\addplot [color=mycolor6, line width=2.5pt, mark=o, mark options={solid, mycolor6}]
  table[row sep=crcr]{%
25000	870.89939912014\\
30000	873.904087122549\\
35000	873.808039642107\\
40000	875.086378889936\\
45000	875.106699719451\\
50000	879.05898946951\\
55000	878.180009495607\\
60000	880.215048691829\\
65000	882.193807249773\\
70000	881.918281364109\\
75000	882.457592893627\\
};
\addlegendentry{BA-SMURF-I}

\end{axis}

\draw[color=gray, very thick](1.8,4.9) ellipse (0.3 and 1.);

\draw[color=gray, very thick](1.8,1.2) ellipse (0.3 and 1.) ;

\filldraw[darkgray] (1.95,5.9) circle (0pt) node[anchor=west]{\shortstack{True\\ Positions}};

\filldraw[darkgray] (1.8,2.6) circle (0pt) node[anchor=west]{\shortstack{Tracked \\ Positions}};

\end{tikzpicture}%
    \vspace{-2\baselineskip}
    \caption{Average throughput obtained by the protocols for different \gls{uav} densities with $M=16$ antenna elements.}
    \label{fig:tracked_densities}
    \vspace{-1.1\baselineskip}
\end{figure}

\begin{figure}
    \centering
\definecolor{mycolor1}{rgb}{0.00000,0.44700,0.74100}%
\definecolor{mycolor2}{rgb}{0.85000,0.32500,0.09800}%
\definecolor{mycolor3}{rgb}{0.92900,0.69400,0.12500}%
\definecolor{mycolor4}{rgb}{0.49400,0.18400,0.55600}%
\definecolor{mycolor5}{rgb}{0.46600,0.67400,0.18800}%
\definecolor{mycolor6}{rgb}{0.30100,0.74500,0.93300}%
\begin{tikzpicture}

\begin{axis}[%
width=0.35\textwidth,
height=2.in,
% width=0.4\textwidth,
% height=2.5in,
at={(0in,0in)},
scale only axis,
xmin=0,
xmax=70,
xlabel style={font=\color{white!15!black}},
xlabel={Number of Antenna Elements},
ymin=740,
ymax=880,
ylabel style={font=\color{white!15!black}},
ylabel={Average Throughput (Mb/s)},
axis background/.style={fill=white},
legend style={at={(1,0.55)}, anchor=east, legend cell align=left, align=left, draw=white!15!black, font=\small}
]
\addplot [color=mycolor1, line width=2.5pt, mark=o, mark options={solid, mycolor1}]
  table[row sep=crcr]{%
1	866.234013844205\\
4	817.435272859977\\
8	798.759103731525\\
16	781.726054236127\\
32	762.245191791073\\
64	743.767858248329\\
};
\addlegendentry{SMURF-T}

\addplot [color=mycolor2, line width=2.5pt, mark=o, mark options={solid, mycolor2}]
  table[row sep=crcr]{%
1	865.46594480557\\
4	864.977881957611\\
8	863.808262880108\\
16	863.629142189301\\
32	864.126152059422\\
64	864.657848017693\\
};
\addlegendentry{SMURF-I}

\addplot [color=mycolor3, line width=2.5pt, mark=o, mark options={solid, mycolor3}]
  table[row sep=crcr]{%
1	863.976700263651\\
4	816.653936950183\\
8	798.068156279175\\
16	781.269874092945\\
32	762.270742098189\\
64	741.77957815169\\
};
\addlegendentry{DBR-T}

\addplot [color=mycolor4, line width=2.5pt, mark=o, mark options={solid, mycolor4}]
  table[row sep=crcr]{%
1	863.86180212462\\
4	862.163779386286\\
8	863.721578130039\\
16	863.810999567334\\
32	863.570115222494\\
64	862.487596654729\\
};
\addlegendentry{DBR-I}

\addplot [color=mycolor5, line width=2.5pt, mark=o, mark options={solid, mycolor5}]
  table[row sep=crcr]{%
1	879.923269693438\\
4	830.911782551045\\
8	813.637405269455\\
16	796.103071034135\\
32	778.531184169929\\
64	757.228604878738\\
};
\addlegendentry{BA-SMURF-T}

\addplot [color=mycolor6, line width=2.5pt, mark=o, mark options={solid, mycolor6}]
  table[row sep=crcr]{%
1	879.388924634316\\
4	876.897733295156\\
8	878.055370111666\\
16	879.05898946951\\
32	878.956005570899\\
64	877.725361842851\\
};
\addlegendentry{BA-SMURF-I}

\end{axis}

\end{tikzpicture}%
    \vspace{-2\baselineskip}
    \caption{Average throughput obtained by the protocols for different \glspl{uav} antenna configurations for a network density of $50000$~UAVs/km$^3$.}
    \label{fig:number_antennas}
    \vspace{-1.\baselineskip}
\end{figure}
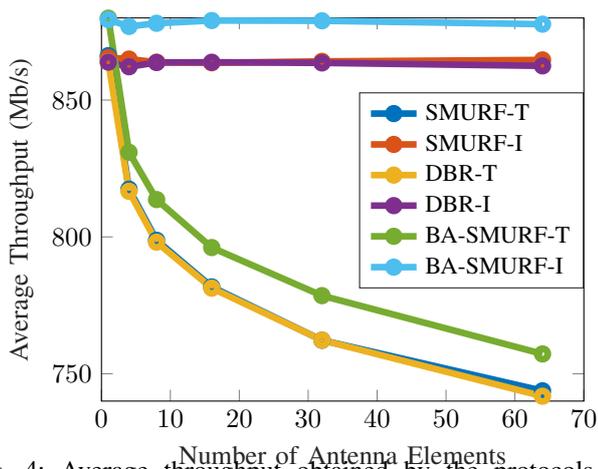
\begin{figure}
    \centering
\definecolor{mycolor1}{rgb}{0.00000,0.44700,0.74100}%
\begin{tikzpicture}

\begin{axis}[%
% width=0.4\textwidth,
% height=2.4in,
width=0.35\textwidth,
height=2.1in,
at={(0in,0in)},
scale only axis,
bar shift auto,
xmin=-0.2,
xmax=7.2,
xtick={1,2,3,4,5,6},
xticklabels={{1},{4},{8},{16},{32},{64}},
xlabel style={font=\color{white!15!black}},
xlabel={Number of Antenna Elements},
ymin=0,
ymax=35,
ylabel style={font=\color{white!15!black}},
ylabel={Average Interference (dB)},
axis background/.style={fill=white}
]
\addplot[ybar, bar width=0.8, fill=mycolor1, draw=black, area legend] table[row sep=crcr] {%
1	32\\
2	9.08327954489335\\
3	5.52889915264825\\
4	3.43644535943081\\
5	2.10048550591383\\
6	1.28874440233995\\
};
\addplot[forget plot, color=white!15!black] table[row sep=crcr] {%
-0.2	0\\
7.2	0\\
};
\end{axis}

\end{tikzpicture}%
    \vspace{-2\baselineskip}
    \caption{Average interference incurred for different \glspl{uav} antenna configurations.}
    \label{fig:interference}
    \vspace{-1.\baselineskip}
\end{figure}
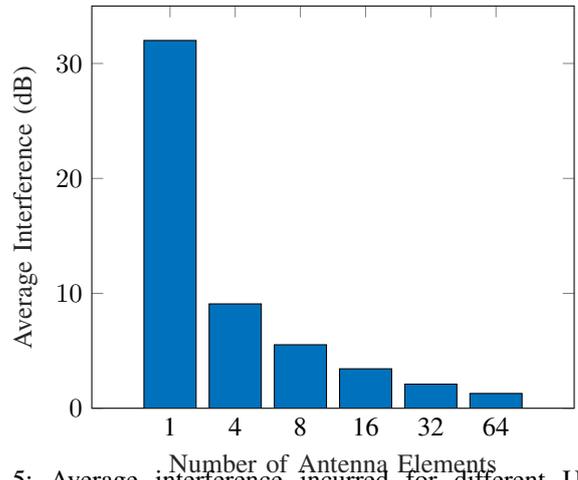

% Rephrase later 
To accurately determine the performance achieved for all the protocols, we carry out simulations with different antenna configurations and different densities. \figurename~\ref{fig:tracked_throughput} compares the average throughput of all the protocols applied in a network with density of $50000$ UAVs/km$^3$ (which corresponds to about 20 drones in network) and for the antenna configuration of $M_H=M_V=4$, i.e., $M=16$ elements in the antenna. Firstly, it is easy to see that the most important factor is the availability of the true position: the real protocols, which operate on uncertain information, have a throughput that is lower by about 10\% than the one obtained by their respective ideal versions. However, \gls{basmurf} outperforms both \gls{smurf} and \gls{dbr} for both true and tracked positions. This indicates that the routes chosen by \gls{basmurf} are able to provide median as well as the worst case throughput (i.e., $25^{\text{th}}$ percentile)  
% This indicates that the routes chosen by \gls{basmurf} are able to provide median as well as $25^{th}$ percentile, i.e., worst case, throughput 
higher than \gls{smurf} and \gls{dbr} for both tracked and true scenarios. Hence, \gls{basmurf} is able to outperform the other two protocols by about 5\% even with a static beamforming scheme.

A similar behavior is observed when comparing all the protocols considering different \gls{uav} densities. Considering the same antenna configuration as above, \figurename~\ref{fig:tracked_densities} illustrates this density comparison. Both \gls{smurf} and the enhanced \gls{basmurf} version can exploit high-density network to find better routes, i.e., routes with high link existence probability (for \gls{smurf}) and high minimum expected capacity (for \gls{basmurf}). On the other hand, \gls{dbr}'s throughput does not increase with increasing density, as the protocol only takes into account the distance between the tracked positions, without considering the uncertainty: consequently, it will not choose safer routes, which are available if the density of the network increases. It is also evident from the figure that \gls{basmurf} outperforms both \gls{smurf} and \gls{dbr} for all densities and for both the tracked and true position information, thanks to its joint consideration of the position uncertainty and beamforming pattern. Additionally, considering the beamforming design can allow the system to potentially reduce interference to other established routes, as well as allowing for more efficient power allocation for the transmission towards the receiver \gls{uav}.

Another interesting evaluation is the impact of different antenna configurations, i.e., different beamforming patterns, assuming a fixed power allocation for the antenna. \figurename~\ref{fig:number_antennas} shows the performance achieved by all the protocols for different antenna configurations in a network with a density of $50000$~UAVs/km$^3$. The performance of the protocols when taking into account true positions of the \glspl{uav} in the network is similar for all different antenna configurations. This shows that  beamforming does not impact the performance when the true positions are known by the protocols, as the transmitter \gls{uav} is able to beamform the signal in the correct direction. However, when considering the tracked position information, beamforming becomes a problem: as the number of elements in the antenna increases, the transmitted beam narrows, leading to a stronger impact of the position uncertainty, and consequently, to a lower throughput. The loss in performance is due to the fact that the central controller does not know the true position of the \glspl{uav} and it devises the beamsteering vector $a_l(\phi, \theta)$ based on the expected position of the \glspl{uav}: the narrower the beam gets, the larger the impact of pointing errors becomes. When the number of elements is 1, the performance of all the protocols is similar for tracked and true position information, since the antenna is omnidirectional. 
Additionally, \gls{basmurf} outperforms \gls{smurf} and \gls{dbr} both when considering perfect information for the protocols and in the more realistic setting for all the antenna configurations. This highlights the importance of incorporating beamforming information to determine routes in \glspl{fanet}.

While the average throughput for the considered route is higher when choosing a single antenna element (i.e., an omnidirectional antenna), the downside of this configuration is the high interference to neighboring \glspl{uav}, as shown in \figurename~\ref{fig:interference}: an omnidirectional transmission will have an increased impact on other transmissions. When the network is interference-limited, the use of narrow beamforming design to reduce the neighbourhood interference is beneficial: the lower received power due to pointing errors is compensated by the lower interference. On the other hand, when the network is noise-limited, the use of wide beamforming is beneficial, as it can reduce the impact of \gls{uav} positioning uncertainty. Our results also show that the need for adaptive beamforming for each \gls{uav} is crucial, as having the same beamforming design for the entire network can reduce the overall performance achieved by the network.

\section{Conclusion}\label{sec:conclusion}
In this work, we provide a statistical analysis of a \gls{fanet} with tracked position information and beamforming design, and derive the minimum expected capacity for both single links and entire routes. We then present the \gls{basmurf} protocol, a multihop routing protocol, which computes the route with the highest expected capacity. The simulation results show that \gls{basmurf} can outperform the existing protocols in realistic network conditions. It is also shown to outperform state of the art approaches for different \gls{uav} network densities and different antenna configurations. 

Moving forward, an interesting avenue of future research is to define beamforming optimization scenarios with respect to the tracking information available for each \gls{uav} in the networks, jointly optimizing them and the routing process based on the \glspl{uav}' tracked positions and the estimation uncertainty.

\bibliographystyle{IEEEtran}
\bibliography{./bibliography.bib}

\end{document}